\documentclass[10pt]{article}
\usepackage[letterpaper]{geometry}
\usepackage{hicss}
\usepackage{times}
\usepackage[none]{hyphenat}
\usepackage{url}
\usepackage{latexsym}
\usepackage{graphicx}
\graphicspath{{images/}}
\usepackage{amssymb}
\usepackage{cite}
\usepackage{tabularx,ragged2e,booktabs,caption}
\usepackage{amsmath,amssymb,amsfonts}
\usepackage{algorithmic}
\usepackage{graphicx}
\usepackage{textcomp}
\usepackage{xcolor,soul}
\usepackage[english]{babel}
\usepackage{amsthm}
\usepackage[english]{babel}
\usepackage{verbatim}
\usepackage{comment}
\usepackage[toc,page]{appendix}

\makeatother
\usepackage[ruled]{algorithm2e}
\usepackage{ulem}
\allowdisplaybreaks

\title{Decentralized Voltage Control with Peer-to-peer Energy Trading in a Distribution Network}

\author{Chen Feng \\
School of Industrial Engineering\\
Purdue University \\
 {\underline{ feng219@purdue.edu}} \\\And
 Andrew L. Lu \\
 School of Industrial Engineering\\
 Purdue University \\
  {\underline{andrewliu@purdue.edu} }\\\And 
 Yihsu Chen \\
 Electrical and Computer Engineering\\
 University of California Santa Cruz \\
 {\underline{yihsuchen@ucsc.edu}} \\
}

\begin{document}
\maketitle
\begin{abstract}
Utilizing distributed renewable and energy storage resources via peer-to-peer (P2P) energy trading has long been touted as a solution to improve energy system’s resilience and sustainability. Consumers and prosumers (those who have energy generation resources), however, do not have expertise to engage in repeated P2P trading, and the zero-marginal costs of renewables present challenges in determining fair market prices. To address these issues, we propose a multi-agent reinforcement learning (MARL) framework to help automate consumers’ bidding and management of their solar PV and energy storage resources, under a specific P2P clearing mechanism that utilizes the so-called supply-demand ratio. In addition, we show how the MARL framework can integrate physical network constraints to realize decentralized voltage control, hence ensuring physical feasibility of the P2P energy trading and paving ways for real-world implementations. 
\end{abstract}
\subsubsection*{Keywords:}

peer-to-peer energy trading, distributed voltage control, multi-agent reinforcement learning



\maketitle
\section{Introduction}\label{sec:Intro}
Distributed energy resources (DERs), such as photovoltaic (PV) panels and energy storage systems, have grown rapidly and have become a key element of a smart energy system. Many potential benefits of DERs include increased use of renewable energy,  making the energy system cleaner and more sustainable, and improved resilience to natural disasters. 
From the perspectives of consumers, such as households and commercial buildings, when equipped with DERs, they become prosumers; that is, they can either buy from or sell energy to the power grid. When compensated properly, consumers will have more incentives to invest in DERs, which, in turn, will accelerate the many benefits of DERs.  

From our perspective, however, there are at least two major challenges to the broad adoption of DERs. How to properly compensate DER owners (aka prosumers) represents one such challenge; the other is how to efficiently utilize the distributed resources to maintain system reliability and feasibility, while not incurring inconvenience or intruding privacy. To address the first challenge, feed-in-tariff (FIT) has been a widely adopted approach to entice DER investment, which is a non-market-based approach that just pays DER owners a fixed rate for the energy they produce. 
However, government agencies or utility companies usually do not know what a fair FIT rate is, and the tariffs are often set too high. As an alternative, peer-to-peer (P2P) market mechanisms have been proposed to integrate and incentivize DERs. Such mechanisms include the bilateral matching approach \cite{P2PWuLei}, double auctions \cite{zhao2021auction}, and the supply-demand-ratio mechanism \cite{liu2017energy}, among others. 
Extensive reviews of such work can be found in \cite{PoorP2PReview,TransactiveEnergyReview}. 

As characterized in \cite{PoorP2PReview}, a P2P energy market consists of two layers: the virtual layer (aka the financial layer) and the physical layer. Many works in the literature focus on one layer only. For example, both \cite{zhao2021auction} and \cite{MARLDA} study decentralized agents in repeated double-auctions in a P2P market, with the former using a multiagent, multi-armed bandit approach, while the later using a multi-agent reinforcement learning (MARL) framework. A similar MARL approach is applied in \cite{Goran2} with the focus on scalability and privacy-preserving of the algorithm. In terms of specific trading and pricing mechanisms in a P2P market, bilateral contract approaches are studied in \cite{morstyn2018bilateral} and \cite{sorin2018consensus}. The former employs a trading mechanism from the economics literature \cite{JPE}; while the later uses a consensus-based algorithm. Other market mechanisms, including bill sharing, mid-market rate and supply-demand ratio (the same mechanism used in this paper), are compared in \cite{MechCompare} using a multi-agent simulation approach. None of the mentioned works consider physical distribution network constraints or voltage control. Such physical-layer considerations are the main focus of \cite{liu2017distributed} and \cite{liu2021online}, among others. 
The former paper uses alternating direction method of multipliers (ADMM) to realize distributed control; while the later one employs a MARL approach. 

Much fewer workers consider both trading mechanisms and physical constraints in a P2P energy market. Some notable works along this direction include the following. \cite{guerrero2018decentralized} employs a matching mechanism similar to stock trading to facilitate bilateral trading among prosumers, which are modeled as zero-intelligence-plus agents. Physical constraints, including voltage, distribution lines and energy losses, are considered through the so-called sensitivity factors; that is, each bilateral transaction is analyzed through their impacts on the network, and infeasible transactions are rejected. In \cite{RLPoor}, network capacity constraints are considered, but the pricing of DER-generated energy is somewhat arbitrary. Two most relevant papers to our work are \cite{MARLDDPG} and \cite{biagioni2021powergridworld}. The former paper uses the supply-demand ratio approach for market clearing. Distribution network constraints are represented by an ACOPF formulation. Prosumers bidding policies, along with the network constraints, are trained through a centralized training, decentralized implementation framework, the so-called multi-agent deep deterministic
policy gradient (MADDPG) algorithm. Such an algorithm requires the pooling of all agents' state and action spaces for training, and can be a major hurdle for real-world implementation. The later paper, \cite{biagioni2021powergridworld}, uses a completely decentralized training-learning framework and imposes system voltage violation on each agent to realize decentralized voltage control. However, their work does not have a specific P2P trading mechanism.

In this work, we propose a framework that combines the strengths of the two papers mentioned above. On the algorithm side, we adopt a completely decentralized MARL approach that does not require centralized learning, and integrate both the virtual and physical layers into the algorithmic framework. At the physical layer,  we explicitly include smart inverters' capability to generate reactive power to aid in decentralized Volt-VAR control. At the virtual layer, we represent each prosumer's bidding problem as a Markov decision process (MDP) and employ the supply-demand ratio mechanism to clear the P2P market. As a result, complete decentralization is realized in both the virtual and physical layers. While having a centralized and independent distribution system operator (DSO) to operate a P2P energy market has been suggested for a long time, there are many issues facing a potential DSO \cite{DSOIssues}, and there are no functional DSOs in the US market. Hence, we aspire that our fully decentralized framework will pave the way for easier participation and better utilization of the vast DER resources.

The rest of the paper is organized as follows. The P2P market and individual prosumer's decision-making problem are presented in Section \ref{sec:MktAgent}. The multi-agent learning framework and the corresponding algorithm are described in Section \ref{sec:MARL}. The numerical simulations and results are presented and analyzed in Section \ref{sec:Num}, followed by the conclusion and future work in Section \ref{sec:Con}.
\section{Problem Formulation}
\label{sec:MktAgent}
In this section, we first describe a clearing mechanism, referred to as the supply-demand-ratio (SDR) mechanism, that can address the two potential issues faced by a renewable energy-dominated P2P market: (i) breakdown of marginal-cost-based pricing due to the all-zero-marginal-cost resources, and (ii) the potential bang-bang outcomes in a double-auction-based mechanism as the reserve prices for buyers (the UR) and the sellers (the FIT) are publicly known. 

Based on the SDR clearing mechanism, we present the formulation of each agent (aka, a prosumer)'s problem as a Markov decision process, which serves as a building block for the MARL framework.   
\subsection{Market Clearing Mechanism}
\label{subsec:market}
Consider a P2P energy market in a distribution network in which consumers and prosumers (all referred to as agents hereinafter) can buy or sell energy. Let $\mathcal{I} = \{1,2,...,I\}$ be the set of all agents in the distribution network. Market clearing is organized in fixed time steps, such as every hour or every 15 minutes, and trade is generally carried out at some time ahead, such as day-ahead or hour-ahead. At each time step $t$, every agent $i \in \mathcal{I}$ can submit bids, denoted by $b_{i,t}$, to buy or sell energy. If $b_{i,t}$ is positive, it represents a bid to sell; while a negative quantity represents a bid to buy. In addition, we use $\mathcal{B}_t$ and $\mathcal{S}_t$ to denote the set of buyers and sellers at a time step $t$, respectively. By the notation above, we have that 
$\mathcal{B}_t = \{i: b_{i,t} < 0\}$ and $\mathcal{S}_t = \{i: b_{i,t} \geq 0\}$. Note that the set of buyers and sellers can certainly change over time, reflecting real-world situations where prosumers may buy or sell energy in a particular time period, depending on their own demand and energy generation.        

To clear the market at time $t$, we adopt the  supply-demand ratio (SDR) mechanism, as proposed in \cite{liu2017energy}. The mechanism is straightforward to describe and to implement. First, the SDR at $t$ is defined as the ratio between total sell and total buy bids; that is,
\begin{align}
	SDR_t := \frac{\displaystyle \sum_{i \in \mathcal{S}_t} b_{i,t}}{\displaystyle -\sum_{i \in \mathcal{B}_t} b_{i,t}}. 
	\label{eq:SDR}
\end{align}
To ensure that $SDR_t$ is well-defined,  we assume that $\mathcal{B}_t \neq \emptyset$ for any $t$.  This is a reasonable assumption as it is highly unlikely in a distribution network that no customer needs energy. (In the extreme case where every customer in the market is a prosumer and they all have excess energy to sell at a time $t$, then the P2P market can simply be suspended and all excess energy sold to a utility or an DER aggregator at a pre-defined price (or the real-time wholesale market price). This will not affect our modeling and algorithm framework in any way.) On the other hand, $\mathcal{S}_t$ can be $\emptyset$ at a given $t$ (meaning that no prosumer has excess energy to sell in $t$). In this case, $SDR_t = 0$. 

With the definition of $SDR_t$, in any particular round $t$, 
if $SDR_t\geq1$, which means there are more sell bids than buy bids, all excess energy in the P2P market is assumed to be sold to a utility company, a DSO or an aggregator at a pre-defined rate, which we just use $FIT$ to denote in general.\footnote{As mentioned in Section \ref{sec:Intro}, the $FIT$ rate is an artificial rate set by utilities/policy makers. For all intents and purposes, it can be set as zero and will not affect the pricing mechanism or our algorithmic framework in any way.}  If $0\leq SDR_t<1$, which means that the sell bids are less than buy bids in round $t$, then the unmet demand bids will purchase energy at a pre-defined utility rate, denoted as $UR$. Without loss of generality, we assume that $FIT < UR$.\footnote{If $FIT > UR$, it means that energy consumers pay more than the utility rate to purchase energy from the prosumers, which is equivalent to direct subsidy from energy consumers (including low-income consumers who do not have the means to install solar panels) to prosumers. This cannot be justified from equity perspective.}

\begin{figure}[!htb]
    \centering
    \includegraphics[scale = 0.37]{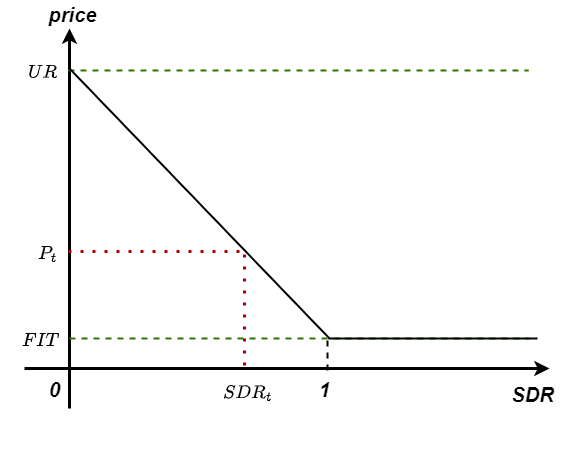}
    \caption{SDR market price function}
    \label{fig:SDR_Price}\vspace*{-5pt}
\end{figure}

Based on the above rules, the market price in a SDR mechanism  can be defined as follows:
\begin{align}
	\label{SDR}
	& P_t:=P(SDR_t) \nonumber \\
	&:= 
	\begin{cases}
		(FIT-UR)\cdot SDR_t + UR, &  0 \leq SDR_t \leq 1\\
		FIT, & SDR_t > 1.
	\end{cases}
\end{align}
$P_t$ is the market clearing price in round $t$, which is a piece-wise linear function that takes the value of $SDR_t$ as input.  as illustrated in Figure \ref{fig:SDR_Price}. 

The interpretation of the $SDR$-based market clearing price is straightforward. When the total supply bids are less than the total demand bids (that is, $0 \leq SDR_t < 1$), the market price is simply a downward-sloped linear function connecting the de facto price ceiling ($UR$) and price floor ($FIT$). 
When the total supply bids exceed total demand bids (that is, $SDR_t > 1$),  the market price is simply the price floor $FIT$. There may be critisms of the $SDR$-approach as the pricing mechanism may seem to be arbitrary (for example, instead of having a linear function, the pricing function could be replaced by any downward-shaped nonlinear functions, which would lead to different market prices with the same $SDR$) and is not based on any economic theories. However,  from our perspectives, the simplicity (and transparency) of the SDR mechansim is well-suited for a P2P energy market with all-zero marginal cost resources. In \cite{zhao2021auction}, it is identified that a local P2P energy market using traditional market clearing mechanisms may not work well due to the special characteristics of such a market. Since uncleared uncleared electricity demand bids have to be met through buying from a utility company at the rate of UR, and uncleared energy supply bids have to be sold to a utility at FIT. Both the UR and the FIT become the de facto price ceiling and price floor (or reservation prices) for all bids, respectively. However, both the UR and the FIT are public information, coupled with the fact that all the supply sources are of zero-marginal costs, making the P2P market outcomes likely of the bang-bang type; that is, when the total demand bids are more than supply bids, the market price would be UR; reversely, the price would be FIT. We view the SDR approach as a fair and reasonable approach to avoid such bang-bang outcomes, especially for a new market in which the supply resources are likely much less than demand. 




Based on the above descriptions, we can also easily write out a prosumer's cost of energy purchase (as a buyer) or profit of energy sales (as a seller) from the P2P market in $t$:
\begin{align}
\label{eq:P2PReward}
	& r^{m}_{i,t} :=
	\begin{cases}
		\mathbb{I}_{i\in\mathcal{B}_t} \times \Big[ SDR_t \cdot P_t \cdot b_{i,t}  \\
		+\  (1-SDR_t) \cdot UR \cdot b_{i,t} \Big]\\
		+ \mathbb{I}_{i\in\mathcal{S}_t} \times \Big(P_t \cdot b_{i,t}\Big) ,& 0 \leq SDR_t \leq 1\\
		FIT\cdot b_{i,t}, & SDR_t > 1,
	\end{cases}
\end{align}
where $\mathbb{I}_{i\in\mathcal{B}_t}$ is an indicator function such that $\mathbb{I}_{i\in\mathcal{B}_t} = 1$ if $i\in \mathcal{B}_t$, and 0 otherwise. Similarly,  $\mathbb{I}_{i\in\mathcal{S}_t} = 1$ if $\in \mathcal{S}_t$ and 0 otherwise. 

\subsection{Markov Decision Process for a Prosumer}
\label{subsec:MDP}
In a local P2P energy market, we assume that each prosumer (such as a household or a commercial building) locates at one of the buses of the distribution network, and has three components under their control: an HVAC system, a photovoltaic (PV) system (either on rooftop or ground-mounted)  with a smart inverter, and an energy storage system. Each prosumer also has fixed base load of energy consumption. In a particular time period $t$, whether a prosumer is a net energy seller or a buyer depends on their energy consumption from base load and HVAC versus the level of PV generation and energy storage. 

Due to the presence of energy storage, each agent's bidding and charging decisions are linked over time, which naturally calls for a dynamic programming framework to model the decision process. As such, 
we formulate an agent's energy trading and control process as a discrete-time MDP, and the key modeling components are described in the following. 
\begin{description}
 \item[$\bullet$] \textit{Observation:} agent $i$ at time step $t$ can observe $s_{i,t} := (h, temp^o_t,p_{i,t},d^p_{i,t}, d^q_{i,t}, v_{i,t}, e_{i,t},P_{t-24})$ from an observation space $\mathcal{S}_i$, where $h$ is the time of a day (such as a specific hour in a day), $temp^o_t$ is the outside temperature, $p_{i,t}$ is the PV (real) power generation, $d^p_{i,t}$ and $d^q_{i,t}$ are the inflexible demand of active and reactive power of agent $i$, respectively, $v_{i,t}$ is the magnitude of the voltage, $e_{i,t}$ is the state of charge of agent $i$'s energy storage system, and $P_{t-24}$ is the past P2P energy market price at the same time on the most recent day (or a random price between $FIT$ and $UR$ if $t - 24 \leq 0$; namely, the first day) observed at time step $t$. We assume that each agent controls only one PV system and one energy storage system. Then each element of the observation is a scalar, and the dimension of the observation space is 8; that is, $\mathcal{S}_i\subset \Re^7$. Note that among the observed states, only $e_{i,t}$ is the state variable affected by agent $i$'s action and state at the previous time step, whose transition function is given below. The other observations, $temp^o_t$ and $p_{i,t}$, are exogenous to agent $i$. It is reasonable to assume that the outside temperature $temp^o_t$ is bounded by the historical highest and lowest temperature, and PV generation $p_{i,t}$ is subject to the PV capacity. Therefore, we can assume that the observation space is bounded. In our simulation, we use simple random samples to create variations of the temperature and PV generation for each agent. The details are given in Section \ref{subsec:Data}. \vspace*{-5pt}
  \item[$\bullet$] \textit{Action/Control variables:} agent $i$ in the time step $t$ takes action $a_{i,t} := (a^{flow}_{i,t}, a^{temp}_{i,t}, a^{q}_{i,t}, a^{s}_{i,t}, b_{i,t})$ from action space $\mathcal{A}_i$, with $a^{flow}_{i,t}, a^{temp}_{i,t}$ being $Z_i$-dimension vectors where $Z_i$ is the total number of zones in agent $i$'s HVAC system, and the other elements being scalars. Hence, the dimension of the action space is $2Z_i+3$; that is, $\mathcal{A}_i\subset \Re^{2Z_i+3}$. The variable $a^{flow}_{i,t}$ is the vector of HVAC flow rate set for each zone of agent $i$'s HVAC system, $a^{temp}_{i,t}$ is the vector of HVAC discharge temperature for each zone of agent $i$'s HVAC system, $a^{q}_{i,t}$ is the reactive power generation by agent $i$'s smart inverter, $a^{s}_{i,t}$ is energy charged (if $a^{s}_{i,t} > 0$) or discharged (if $a^{s}_{i,t} < 0$) of the battery, and $b_{i,t}$ is agent $i$'s energy bid to the P2P market in time $t$ (either buy or sell). The reactive power generation and energy charge or discharge are subject to PV and battery capacity, respectively. It is also reasonable to assume that there are lower and upper bounds for the control of HVAC flow rate and discharge temperature. Therefore, we can assume that the action space $\mathcal{A}_i$ is bounded.
  \item[$\bullet$] \textit{State transition:}  The state transition function for agent $i$ is:
  \vspace*{-15pt}
  \begin{align}
     \begin{split}
      e_{i,t+1} = & \max \Big\{ \min \Big(e_{i,t} + \eta^c_{i} \max(a^{s}_{i,t},0) \\
      &+ \frac{1}{\eta^d_{i}} \min(a^{s}_{i,t},0), \overline{e}_i\Big), 0\Big\},
       \end{split}\vspace*{-5pt}
  \end{align}
  where $\eta^c_{i}$ is the charging efficiency for agent $i$ defined by the ratio of energy charged to the battery divided by the energy consumed in charging, $\eta^d_{i}$ is the discharging efficiency for agent $i$ defined by the ratio of energy output from the battery divided by the energy consumed in discharging, and $\overline{e}_i$ is the battery capacity of agent $i$. \vspace*{-5pt}
   \item[$\bullet$] \textit{Reward:} 
   Each agent $i$'s reward function has three components as follows:
     \begin{align}
      r_{i,t} := r^{c}_{i,t} + r^{m}_{i,t} + r^{v}_t/I.
      \label{eq:reward}
  \end{align}
The first term, $r^{c}_{i,t}$, is the reward for the thermal comfort of agent $i$' from HVAC system, which follows the reward model in the NREL project \cite{biagioni2021powergridworld}:
  \begin{align}
 \begin{split}
           r^{c}_{i,t}: = &-\sum_{j=1}^{Z_i}\Big[\max (0, temp^{j}_{i,t}-\overline{temp})^2\\
           &+\max (0, \underline{temp}-temp^{j}_{i,t})^2 \Big],
 \end{split}
  \end{align}
  where $temp^{j}_{i,t}$ is the temperature of zone $j$ of agent $i$ at time step $t$, which is determined by the outside temperature, the flow rate of zone $j$, and the discharge temperature of zone $j$ of agent $i$'s HVAC system.\footnote{\label{note1}Details of the HVAC system model can be found at \protect\url{https://github.com/NREL/PowerGridworld}}  The two parameters, $\overline{temp}$ and $\underline{temp}$, represent the upper and lower limits of the comfort temperature range, respectively. 
  
  The second term $r^{m}_{i,t}$ is the energy purchase cost or sales profit of agent $i$ from the P2P energy market, as described in Equation (\ref{eq:P2PReward}) in Section \ref{subsec:market}. 
  
  The third term, $r^{v}_t$, is the system voltage violation penalty defined as in \cite{biagioni2021powergridworld}:
  \begin{align}
   \begin{split}
      r^{v}_t:= &-\lambda \sum_{j:Bus}\Big[\max (0, v^j_{t+1}-\bar{v})\\
      &+\max (0, \underline{v}-v^j_{t+1})\Big], 
 \end{split}
 \label{eq:penalty}
  \end{align}
  where $v^j_{t+1}$ is the voltage magnitude of Bus $j$ after each agent makes the decision, and $\bar{v},\underline{v}$ is the upper and lower limit of the bus voltage. $I$ is the total number of agents in the system, and $\lambda$ is a sufficiently large number. We assume that the violation penalty is shared equally by all agents in the system. To calculate $v^j_{t}$ at each bus, after all supply and demand bids have been cleared in $t$, we solve the standard bus-injection model as follows:
  \begin{align}
  \begin{split}
      p_{k}=&\sum_{j=1}^{N}|V_{k}||V_{j}|(G_{k j} \cos (\theta_{k}-\theta_{j}) \\
      &+B_{k j} \sin (\theta_{k}-\theta_{j})), \\
q_{k}=&\sum_{j=1}^{N}|V_{k}||V_{j}|(G_{k j} \sin (\theta_{k}-\theta_{j})\\
&-B_{k j} \cos (\theta_{k}-\theta_{j})) \\
&\text{ for } k= 1,2,\dots, N,
  \end{split}
  \label{eq:BusInjection}
\end{align}
where $N$ is the number of buses in the network, $j$ and $k$ are bus subscripts, $p_k$ is net real power injection at bus $k$, $q_k$ is the net reactive power flow at bus $k$, $G_{k j}$ and $B_{k j}$ are the real and imaginary parts of the admittance of branch $kj$ respectively, $V_{k}$ is the voltage of bus $k$, and $\theta_{k}$ is the phase angle at bus $k$.

After market clearing and solving for voltage magnitudes, the total reward for the agent $i$ at time step $t$ is then realized. Note that while we do not explicitly assume to have a DSO in the distribution network, there still has to be an entity to solve Equation \eqref{eq:BusInjection}. A utility company can certainly assume such a role. The solution process may even be carried out by a distributed ledger system, such as on a blockchain. We want to emphasize that since the market clearing happens before real-time delivery, the voltage regulation penalty is not real financial penalty; it is only some information to be sent to each agent's reinforcement learning algorithm for training and learning purposes. 
  \item[$\bullet$] \textit{Objective function:} At each time step $t$, agent $i$ observes the environment (temperature and PV generation) and its state of charge in the battery $e_{i,t}$, decides on an action $a_{i,t} \in \mathcal{A}_i$, and then receives a reward $r_{i,t}$. A policy $\pi_i$ in this case is a conditional joint probability density function of actions in  $\mathcal{A}_i$, given an observation in $\mathcal{S}_i$;
  that is, agent $i$'s choice of $a_{i,t}$ at time $t$ given an observation $s_{i,t} \in \mathcal{S}_i$ follows the joint density function $\pi_i(\cdot|s_{i,t}): \mathcal{A}_i \to [0,\infty)$. The expected long-term payoff of agent $i$ under policy $\pi_i$ is defined as:
  \begin{align}
  \label{payoff}
      J_i(\pi_i) = \mathbb{E}_{\pi_i}[\sum_{t=0}^{\infty} \gamma_i^t r_{i,t}], 
  \end{align}
  where $\gamma_i \in (0,1)$ is agent i's discount factor. 
  
  Note that the temperature is assumed to be finite, the voltage is based on a per-unit system and hence bounded, and the reward function on the action and observation space is bounded. Therefore, $J_i$ is finite and there exists an optimal policy $\pi_i^*$ such that $\pi_i^* = \sup_{\pi_i} J_i(\pi_i)$. Each agent employs certain algorithms to try to find such an optimal policy.
  
\end{description}

\section{Multi-agent Reinforcement Learning}
\label{sec:MARL}
While the previous section describes the individual agent's decision-making problem as an MDP, the agents' rewards (and long-term payoff) depend on the collective actions of all the agents. Before introducing the multi-agent framework, we first discuss algorithms for a single agent to solve their MDP problem.  

As described in Section \ref{subsec:MDP}, each agent has only partial observation of the system (that is, an agent does not observe other agents' state variables). In addition, each agent's reward is determined by a complex dynamic HVAC model which may not be known in real life, and it depends on the solution to a non-linear power flow equation system which is hard to solve for large scale. Therefore, it is unrealistic to figure out the value function and optimal policy from the state-transition and reward function in our case by using a model-based reinforcement learning (RL) approach, including classic MDP or general dynamic programming algorithms. In contrast, model-free algorithms, such as actor-critic reinforcement learning algorithms, do not need the information about state-transition and reward function when solving the MDP problem. In general, such algorithms learn the value function and optimal policy by trying different policies many times and measuring their rewards. It has been shown to perform well when dealing with problems where state-transition and reward function are complex or inaccessible \cite{konda1999actor}.  Therefore, the actor-critic reinforcement learning algorithm can be a good approach to our problem. To start with, first define the state-action value function, or Q-function under policy $\pi$ for agent $i$, $Q^{\pi}(s_i, a_i)$ as the expected long-term payoff of taking action of $a_i$ in state $s_i$, and following policy $\pi$ afterward:
\vspace*{-8pt}
\begin{align}
    Q_i^{\pi}(s_i, a_i) = \mathbb{E}_{\pi} \Big[\sum_{t=0}^{\infty} \gamma^t r_{i,t} | s_{i,0} = s_i, a_{i,0} = a_i \Big].
\end{align}
Then define the state value function of a state $s$ under policy $\pi_i$ for agent $i$ as:
\vspace*{-8pt}
\begin{align}
    V_i^{\pi_i}(s_i) = \mathbb{E}_{\pi}[\sum_{t = 0}^{\infty}\gamma^t r_{i,t}|s_{i,0} = s_i].
\end{align}

In an actor-critic algorithm, for agent $i$, ``actor" is a function with parameter $\theta_i$: $\pi_{\theta_i}: \mathcal{S}_i \times \mathcal{A}_i \to [0,\infty)$  that characterizes the agent's policy $\pi_i$, and ``critic" is a function with parameter $\omega_i$: $V_{\omega_i}: \mathcal{S}_i \to \mathcal{R}$ that approximates agent $i$'s state value function under policy $\pi_i$. In most of the state-of-the-art actor-critic algorithms such as proximal policy optimization (PPO)\cite{schulman2017proximal}, both the ``actor" and ``critic" are represented by neural networks. Algorithm 1 shows the pseudocode for the training of actor and critic networks using PPO approach \cite{schulman2017proximal}.

\begin{figure*}[!h]
    \centering
    \includegraphics[scale = 0.13]{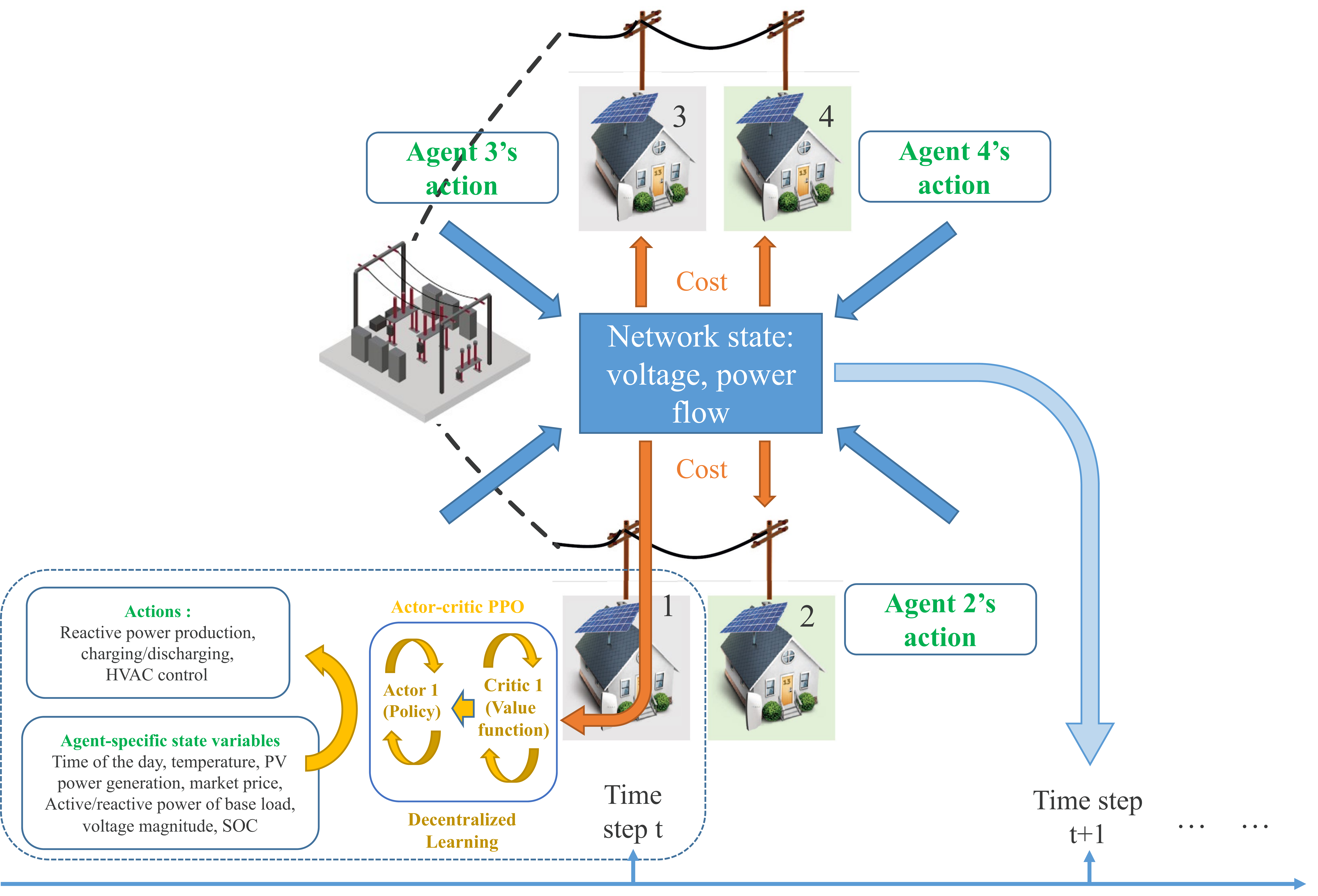}
    \caption{Architecture of the MARL framework}
    \label{fig:MARL}
\end{figure*}

In this work, the traditional single-agent RL framework is extended to a decentralized multi-agent reinforcement learning framework, in which agent $i$ still learns an individual actor network $\pi_{\theta_i}$ and critic network $V_{\omega_i}^{\pi_{\theta_i}}$  based on the agent's own observations and actions. 
This is illustrated in Figure \ref{fig:MARL}.

Compared to MARL algorithms for centralized training and decentralized execution, such as the MADDPG algorithm \cite{lowe2017multi}, our fully decentralized approach has the advantage of less computational burden and greater protection of agents' privacy. The drawback of fully decentralized MARL is that it may have poor performance due to the non-stationary environment from the perspective of each single agent. However, in our case, agents are coupled through the P2P market and system voltage in their reward functions. Each agent is able to partially observe the market and voltage information via the historical prices and system voltage violation penalty; hence the impact of non-stationarity of the state variables on each agent's decision-making can be reduced. Therefore, the fully decentralized MARL, rather than centralized training and decentralized execution MARL approach, is chosen for our case. Specifically, each agent trains an individual PPO policy separately, as in Algorithm \ref{algorithm 2} (the agent index $i$ is omitted). 

\begin{algorithm}
Initialize training dataset $\mathcal{D}$, critic and actor parameters $\theta$ and $\omega$
($\theta$ and $\omega$ are initialized by Glorot uniform initializer as in \cite{glorot2010understanding});\\
\For{\textbf{each} time step t = 0,1,...,T}
{
$\text{1: observe} \ s_{t}$ and \text{adopt} $a_{,t} \sim \pi_{\theta_i}(\cdot|s_{t});$\\
2: The power flow equation is solved and
$r_t$ is computed according to Eq. \eqref{eq:penalty} ; $(s_{t}, a_{t}, r_t, s_{t+1})$ is stored in  $\mathcal{D}$; \\ 
3: Update policy parameters $\theta$ by solving the problem: \\
$\max _{\theta}\frac{1}{|\mathcal{D}|} \sum_{\tau \in \mathcal{D}} \min \Big(M, g(\epsilon, A^{\pi_{\theta}}(s^{\tau}, a^{\tau}))\Big)$
via stochastic gradient ascent, where 
$$M = \frac{\pi_{\theta}(a^{\tau} \mid s^{\tau})}{\pi_{\theta}(a^{\tau} \mid s^{\tau})} A^{\pi_{\theta}}(s^{\tau}, a^{\tau}),$$
$$g(\epsilon, A)= \begin{cases}(1+\epsilon) A & A \geq 0 \\ (1-\epsilon) A & A<0\end{cases},$$
$\epsilon$ is a hyperparameter, say, $\epsilon = 0.2$, and
$A^{\pi}(s, a):=Q^{\pi}(s, a)-V^{\pi}(s)$;\\
4: Fit value function parameters by regression using mean-squared error:  $\omega^*=\arg \min _{\omega} \frac{1}{|\mathcal{D}|} \sum_{\tau \in \mathcal{D}} (V_{\omega}(s^{\tau})-r^{\tau})^{2}$;
}

\caption{PPO algorithm with distributed voltage control in a P2P market for each agent}\label{algorithm 2}
\end{algorithm}

\section{Simulation and Numerical Results}
\label{sec:Num}
\subsection{Test Case} 
\label{subsec:Data}
\begin{figure}[!h]
    \centering
    \includegraphics[scale = 0.5]{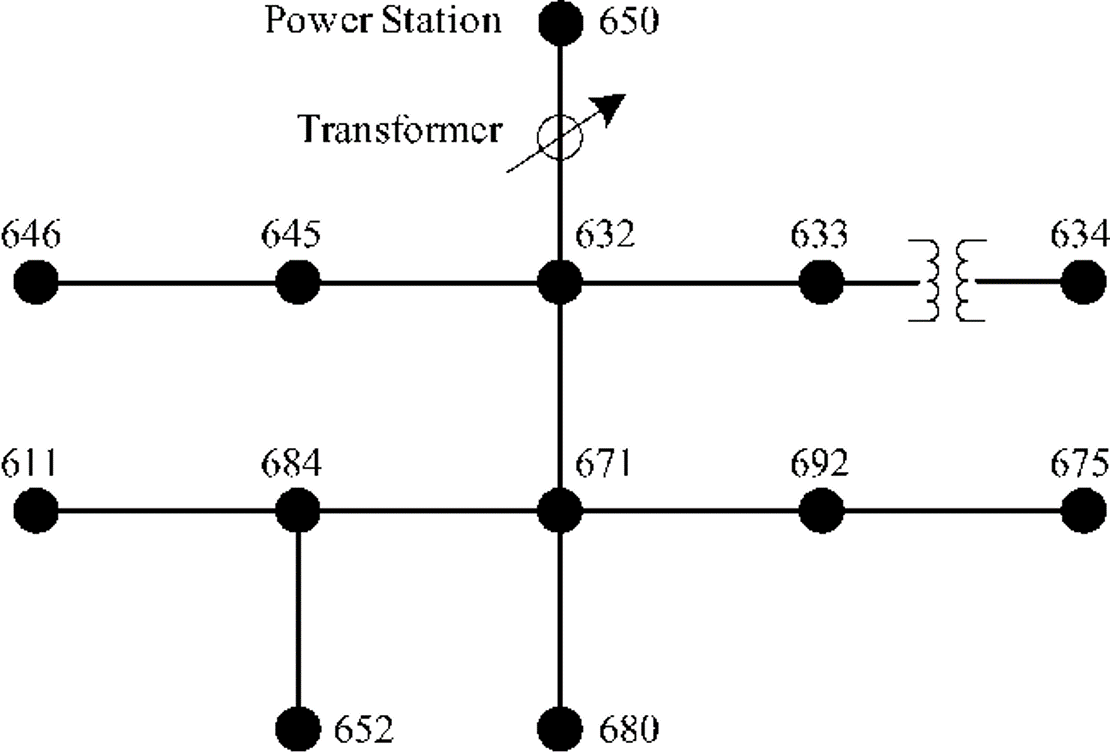}
    \caption{IEEE-13 test feeder}
    \label{fig:feeder}
\end{figure}

We test our problem using the IEEE 13-bus test case. The topology of the test network is shown in Figure \ref{fig:feeder}. 
Detail information on the test feeder, such as the base load of each bus and line data, can be found in \cite{saha2016modeling}. In our case, each bus has one prosumer, except for the substation bus; hence, there are 12 prosumers (aka agents) in the test case. 

\begin{figure}[!h]
  \centering
    \includegraphics[scale = 0.35]{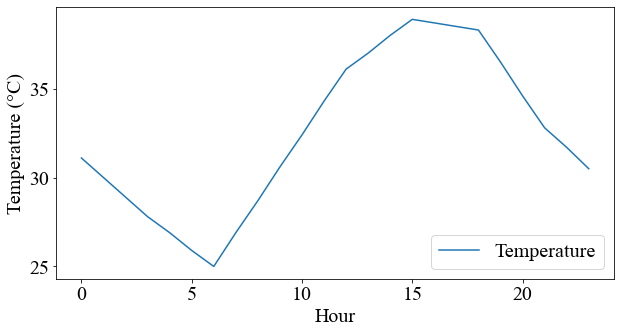}
    \captionof{figure}{Average outside temperature over one day}\label{fig:temp}
\end{figure}

\begin{figure}[!h]
  \centering
    \includegraphics[scale = 0.445]{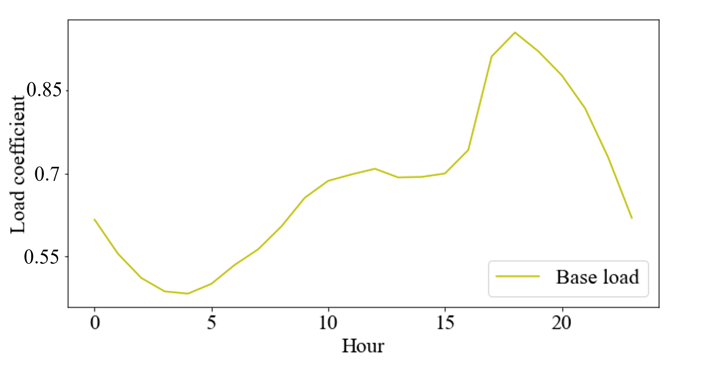}
    \captionof{figure}{Average base load shape over one day (the shape is the same for all buses). The load coefficient of a bus in hour $t$ is the ratio of hour $t$'s base load to the base load of the corresponding bus in IEEE 13  (\protect\url{https://github.com/tshort/OpenDSS/tree/master/Distrib/IEEETestCases/13Bus}).}\label{fig:load}
\end{figure}

\begin{figure}[!h]
  \centering
    \includegraphics[scale = 0.6]{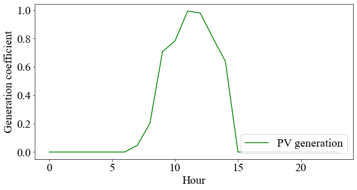}
    \captionof{figure}{Average PV generation shape over 1 day. The generation coefficient of Hour $h$ is the ratio of PV real power generation of Hour $h$ to the PV real power capacity, which is assumed to be  $30 kW$ for each agent.}\label{fig:PV}
\end{figure}

The time step is set to be one hour.  
The baseload ($d_{i,t}$) and PV generation ($p_{i,t}$) of each prosumer, and the outside temperature ($temp^o_t$) are all random variables. We assume that the expected values of these random variables follow fixed daily shapes, which are the same for all prosumers. More specifically, the shape of the average daily temperature is shown in Figure \ref{fig:temp}. 
The average daily base load and PV generation shapes are represented in the form of percentage to their peak values, instead of absolute values, as shown in Figure \ref{fig:PV} and Figure \ref{fig:load}, respectively. 

To generate specific observations by agent $i$ at time $t$, a random sample based on a uniform distribution is drawn at each $t$, with the mean being the values indicated in Figures \ref{fig:temp}, \ref{fig:load} and \ref{fig:PV}, and the range is 0.95 to 1.05 times of the mean values. 

The energy storage setting is assumed to be the same for all prosumers, with the storage capacity set at $50 kWh$, and the charging and discharging efficiency being 0.95 and 0.9, respectively. The output of the smart inverter and PV panel is contrained by a apparent power of 50 VA. The initial state of storage can be arbitrary; we just set it to zero for each agent at the very beginning. Assume that each agent has the same 5-zone HVAC system. The distribution network's voltage limit is set to $[0.96 \ pu,1.04 \ pu]$. The lower and upper bound of the comfort temperature range for each prosumer is set at 22$^{\circ}$C and 28$^{\circ}$C, respectively. $\lambda$ is set to $10^4$ and $\gamma$ is set to 0.999 for all agents. $UR$ and $FIT$ are set at 14 cents / kWh and 5 cents / kWh for all time, respectively.

Policy training is implemented in RLLib, \footnote{Details of RLLib can be found at \protect\url{https://docs.ray.io/en/latest/rllib/index.html}} which is a platform for reinforcement learning algorithm training built on Python. It allows users to create custom ``environment" that can describe their MDP problems on the platform.  Furthermore, in the custom ``environment", users can train existing RL algorithms in RLLib or self-developed algorithms in TensorFlow or PyTorch.  To create an environment for our own problem, we add a P2P market and VVC features to the power system environment built by NREL \cite{biagioni2021powergridworld}\footnote{Details of the package can be found at \protect\url{https://github.com/NREL/PowerGridworld}} on RLLib. The power flow equations are solved by the open distribution system simulator (OpenDSS) in Python.\footnote{\label{note3}Details of OpenDSS package in Python can be found at \protect\url{https://pypi.org/project/OpenDSSDirect.py/}} The algorithm used to train the policy is the PPO built in RLLib.

\subsection{Result Analysis}
\label{sec:ResultAnalysis}
\begin{figure}[h]
    \centering
    \includegraphics[scale = 0.35]{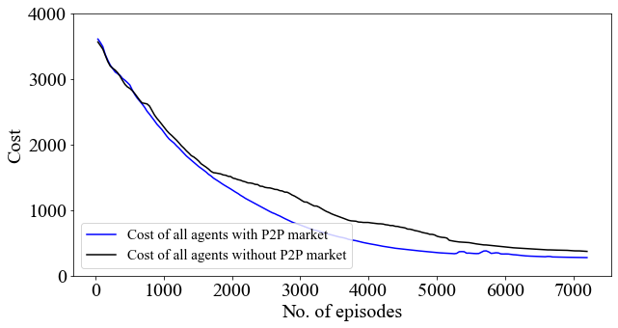}
    \caption{30-episode moving average of episodic total cost}
    \label{fig:Reward}
\end{figure}

\begin{figure}[h]
    \centering
    \includegraphics[scale = 0.48]{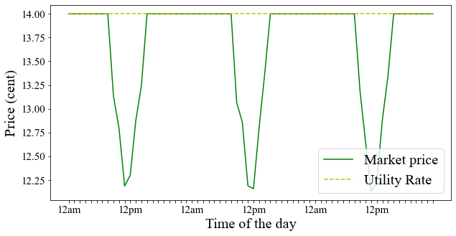}
    \caption{Hourly prices over the last three days}
    \label{fig:price}
\end{figure}

\begin{figure}[h]
    \centering
    \includegraphics[scale = 0.33]{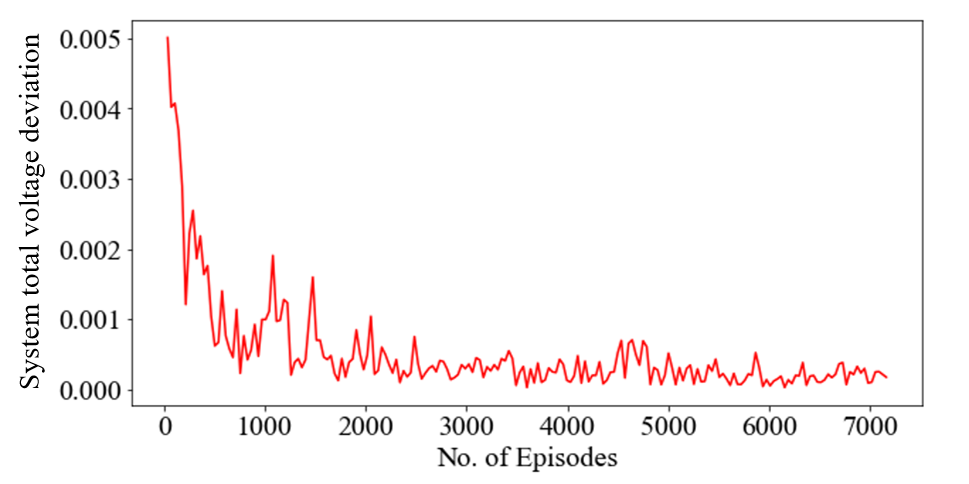}
    \caption{System voltage deviation. The total system voltage deviation (p.u.) in the $n$-th episode is calculated as $\sum_{t=24(n-1)}^{24n-1} \sum_{j:Bus}[\max (0, v^j_{t}-\bar{v})+\max (0, \underline{v}-v^j_{t})]$.}
    \label{fig:violation}
\end{figure}

To present the numerical results, we first introduce the term episodic total cost. The episodic total cost of the $n$-th episode is defined as the total discounted cost of all agents over the $n$-th episode (24 hours): $ \sum_{t=24(n-1)}^{24n-1} \sum_{i=1}^{12} \gamma^t r_{i,t}$. The convergence curves of the 30-episode moving average of the episodic total cost in the distribution network with and without a P2P energy market are shown in Figure \ref{fig:Reward}. In the case without P2P markets, agents can only buy energy from the grid at the utility rate and sell energy to the grid at FIT. Agents still use the distributed PPO algorithm to determine their buying/selling and charging strategies.
For the purpose of comparison, we also present the results for the case without a P2P market. In such a case, the sellers receive payments at $FIT$ and the buyers pay the utility rate $UR$.
It can be seen that the mean episodic total costs gradually converge to a low level at the end of the training in both cases, which shows that the MARL decentralized training approach performs well in terms of convergence in our case. Furthermore, agents in the distribution network with a P2P energy market reach a lower stationary cost level than those without a P2P energy market. This is expected as the introduction of the P2P market allows agents to purchase energy at a price lower than $UR$ and sell at a price higher than $FIT$.

Figure \ref{fig:price} presents the market price of the P2P market in chronological order over the last three days after thousands of training episodes. 
The market price remains at $UR$ one third of the day and drops from it only around noon on each day. This is reasonable because the peak hours of PV generation are around 12 pm, as shown in Figure \ref{fig:PV}. The market price remains at $UR$ during the rest of the day, as energy storage and PV generation cannot meet the demand for each prosumer and hence, no supply bids into the P2P market. 

Figure \ref{fig:violation} shows the gradual reduction of the system voltage deviation after training. It can be seen that the voltage violation of the system is high at the beginning. However, it drops quickly and converges to almost zero after half of the training process. On the one hand, the results are remarkable in the sense that voltage violation reduction is realized through purely decentralized learning. As specified in Eq. \eqref{eq:penalty}, each agent will receive a penalty if the collective bids submitted by all agents lead to a voltage violation. The agents then simply try to submit a different bid. This is entirely a trial-and-error process (with the underlying assumption that a feasible collection of bids exists, and hence can be eventually reached). This process is possible since a bid round always takes place some time before the physical delivery of energy. On the other hand, this trial-and-error approach to reduce voltage violation may be viewed as inefficient, and it is hard to establish any theoretical results to guarantee the convergence (to a steady state) of the decentralized learning process. 
An immediate next step of this research is to investigate if there is any (privacy conserving) communication mechanism among agents that would help accelerate the voltage violation reduction, as well as leading to provable convergence of the overall MARL framework. Two possible general approaches along this line have been proposed in \cite{MARLBasar,MARLAnt}. Nevertheless, we hope that our current work can serve as a benchmark to be compared with other decentralized MARL algorithms that integrate market clearing and physical constraints.

\section{Conclusion and Future Work}
\label{sec:Con}
In this work, we introduced a decentralized framework to integrate DERs, including energy storage, in a market-based approach while maintaining the feasibility of a distribution system. Within such a framework, 
each agent's decision problem in a repeated P2P market can be automated through an RL-based algorithm, as the learning and training of the algorithms are completely decentralized. In addition, the privacy of the agents can be preserved, as no exchange of private information is needed. 
The initial numerical simulation shows encouraging results, as this framework is able to achieve lower energy costs for all prosumers and realize decentralized voltage regulation for the distribution network upon convergence. 

Future work will focus on two directions. First, as mentioned earlier, whether there exists a provably convergent MARL algorithm that is applicable to a distribution-level energy market is an open question. We are actively investigating in this direction. Second, the scalability of our framework (or any MARL-based approach) to the size of real-world systems still needs to be tested. One potential approach to improve the performance and scalability is to study the possibility of increasing communication between neighbors without disclosure of privacy to improve the performance or our MARL framework. 

\bibliographystyle{apalike}
\bibliography{hicss_final}

\end{document}